\documentclass[aps,prl,groupedaddress,twocolumn]{revtex4-1}

\usepackage{amsmath,amssymb}

\usepackage{eqnarray}
\usepackage[colorlinks]{hyperref}
\usepackage{epstopdf} %converting to PDF
\usepackage{color}
\usepackage{lipsum}
\usepackage{graphicx}% Include figure files
\usepackage{dcolumn}% Align table columns on decimal point
\usepackage{bm}% bold math
\newcommand{\be}{\begin{equation}}
\newcommand{\ee}{\end{equation}}
\newcommand{\la}{\langle}
\newcommand{\ra}{\rangle}
\newcommand{\bx}{{\bf{x}}}

\newcommand{\bX}{{\bf{X}}}
\newcommand{\obx}{\overline{\bf{x}}}

\newcommand{\R}{{\rm{I\!R}}}

\renewcommand{\leq}{\leqslant}
\renewcommand{\geq}{\geqslant}

\newcommand{\domain}{\mathcal{S}}

\newcommand{\IO}{\mathcal{I}_{\Omega}}

\begin{document}

\title{Seeding Dispersal Modeling For Systems of Planar Microbial Biofilms}% Force line breaks with \\
%%%%%%%%%%%%%%\thanks{A footnote to the article title}%

\author{Andrea Trucchia$^{1,2}$}
%\email{atrucchia@bcamath.org}
\author{Federica Villa$^{3}$}
\author{Luigi Frunzo$^{4}$}
\author{Gianni Pagnini$^{1,5}$}
\email{gpagnini@bcamath.org}

% \altaffiliation[Also at ]{Physics Department, XYZ University.}%Lines break automatically or can be forced with \\

\affiliation{%
$^1$BCAM  - Basque Center for Applied Mathematics, Alameda de Mazarredo 14, E-48009  
Bilbao, Basque Country -- Spain
}%

\affiliation{%
$^2$Universidad del Pa\'is Vasco/Euskal Herriko Unibertsitatea UPV/EHU, Campus de Leioa, E-48949 Leioa, 
Basque Country -- Spain
}%

\affiliation{%
$^3$Department of Food, Environmental and Nutritional Sciences, University of Milan, 
via Celoria 2, I-20133 Milan,Italy
}%

\affiliation{%
$^4$Department of Mathematics and Applications "Renato Caccioppoli",
University of Naples "Federico II", via Cintia, Monte S. Angelo, I-80126 Naples, Italy
%$^4$Dipartimento di Matematica e Applicazioni
%"Renato Caccioppoli" Universit\`a degli Studi di Napoli Federico II, 
}%

\affiliation{%
$^5$Ikerbasque - Basque Foundation for Science, Calle de Mar\'ia D\'i­az de Haro 3, E-48013 Bilbao, 
Basque Country -- Spain
}%

%\date{\today}% It is always \today, today,
             %  but any date may be explicitly specified

\begin{abstract}
%We propose a modelling approach for the growth of mono-layer microbial biofilm on inert surfaces 
%by focusing on the biofilm spread induced by dispersal, motivated by the lack in the understanding of 
%biofilm dispersal due to the fact that fate and behavior of dispersal cells cannot, as yet, 
%be predicted from the genome. 
%Supported by experimental qualitative comparison, the model results to be promising and allowing 
%for further improvements through more detailed sub-models for front propagation, seeding, 
We propose a modeling approach to study how mature biofilms spread and colonize new surfaces by predicting the formation and growth of satellite colonies generated by dispersing biofilms.
This model provides the basis for better understanding the fate and behavior of dispersal cells, phenomenon that cannot, as yet, be predicted from knowledge of the genome.
%The objective was to have a better understanding of biofilm dispersal because its behavior cannot be predicted from the genome.
The model results were promising as supported by the experimental results. The proposed  approach allows for further improvements through more detailed sub-models for front propagation, seeding, 
availability and depletion of resources. The present study was a successful proof-of-concept in answering the following questions: Can
 we predict the colonization of new sites following biofilm dispersal?
Can we generate patterns in space and time to shed light on seeding dispersal? 
That are fundamental issues for developing novel approaches to manipulate biofilm formation 
in industrial, environmental and medical applications.

%An article usually includes an abstract, a concise summary of the work
%covered at length in the main body of the article.
%\begin{description}
%\item[Usage]
%Secondary publications and information retrieval purposes.
%\item[PACS numbers]
%May be entered using the \verb+\pacs{#1}+ command.
%\item[Structure]
%You may use the \texttt{description} environment to structure your abstract;
%use the optional argument of the \verb+\item+ command to give the category of each item.
%\end{description}
\end{abstract}

%\pacs{Valid PACS appear here}% PACS, the Physics and AstronomyS
                             % Classification Scheme.
%\keywords{Suggested keywords}%Use showkeys class option if keyword
                              %display desired
\maketitle

%\tableofcontents

%%%PEZZO SOSTITUITO DA NOTA n.1 DI FEDERICA!%%% Microbial biofilms are communities of bacterial cells encased in a self-produced polysaccharide matrix and attached to an inert or living surface \citep{alexander-cjb-1982}te{}.%%%%

\begin{figure}
\includegraphics[width=0.4\textwidth , trim={0.5cm 2.5cm 0.5cm 2.5cm},clip]{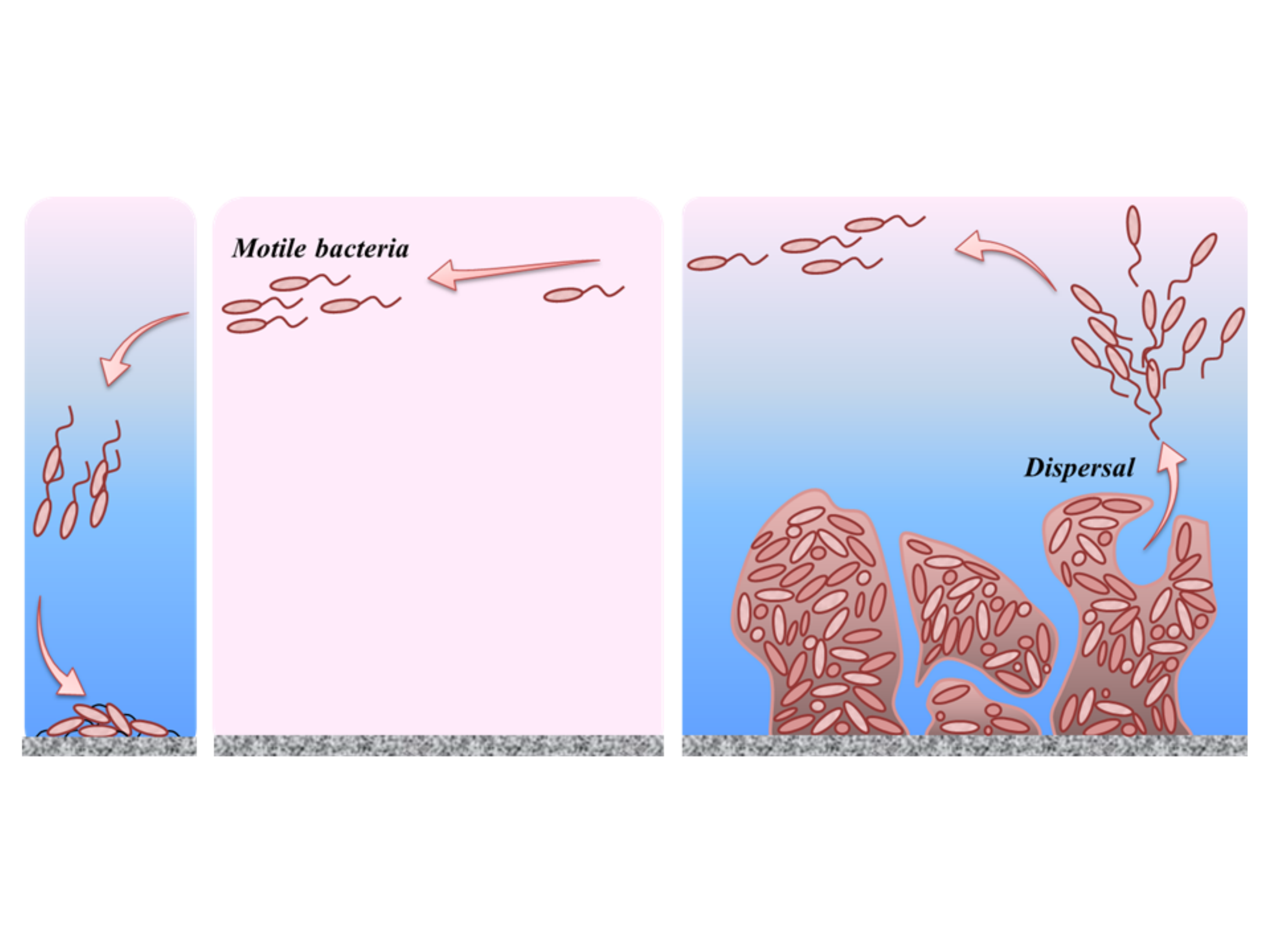} 
\caption{Graphic picture of biofilm seeding mechanism, 
with motile bacteria abandoning the main colony in order to attach into favorable spots where to start new colonies.}
\label{Fig1}
\end{figure}

It is now well accepted that microorganisms lead social lives and engage in complex behavior in response to other organisms and the extracellular environment.
 By adopting coordinated chemical and physical interactions, microorganisms establish complex communities attached to a surface and embedded in
  a self-produced extracellular polymeric matrix, enabling cells to develop efficient survival strategies
   \cite{nadell_etal-nrm-2016}. This sessile lifestyle is called biofilm, and it represents the dominant mode of microbial life
    in many natural, medical and engineered systems \cite{hallstoodley_etal-nrm-2004,mattei_etal-jmb-2017}.
Cells in biofilms undergo developmental programs resulting in an ordered and predictable transition through
 a series of stages, each based on stage-specific expression of genes \cite{flemming_etal-nrm-2016}. The biofilm developmental program culminates
  with the release of free-living cells that can colonize new habitats, possibly richer in resources \cite{mcdougald_etal-nrm-2011}, as seen  Fig. \ref{Fig1}.
%%%% QUI HO AGGIUNTO IL possibly richer in resources, sotto si spiega il motivo. Andrea Trucchia %%%%%%%%%%%
 While detachment is a passive process of cell loss resulting from sloughing of cells and erosion from the biofilm, active or seeding dispersal
  is coordinated via regulatory systems in response to a number of cues (e.g., alteration in the availability of nutrients,
   oxygen depletion, levels of iron) and signals (e.g., acyl-homoserine lactones, diffusible fatty acids, cell-cell autoinducing peptides)  \cite{guilhen_etal-mm-2017}. Thus, seeding dispersal can occur in the complete absence of flowing conditions, and does not depend upon shear forces that removes cells from the biofilm. Another interesting feature of seeding dispersal is that cells appear to have a distinct phenotypes different from those of  biofilm and planktonic cells, increasing cell ability to colonize a greater range of habitats important for niche expansion \cite{chua_etal-nc-2014,dacunto_etal-mb-2015}. 
Thus, dispersal represents an important adaptive strategy with profound impacts on the survival and fitness of microorganisms. It allows biofilm populations to spread and colonize new surfaces, avoiding overcrowding, depletion of resources and competition among cells in the local environment, and promoting the rejuvenation of biofilms  \cite{barraud_etal-ms-2015}. Furthermore, dispersal is linked to the generation or maintenance of genetic variation, with significant outcomes for the success of those bacteria in the environment \cite{purevdorj_etal-m-2005,chua_etal-nc-2014,dacunto_etal-mb-2015}. 
Although dispersal is advantageous from the microbial standpoint, it may negatively affect some industrial and medical processes.
 For instance, through dispersed cells, biofilm can spark new infections within the host and result in the transmission of bacteria between
  different hosts \cite{koo_etal-nrm-2017}. Furthermore, dispersal may promote, for example, the spread of parasitism phenomena
   in animals and plants  \cite{villa_etal-fm-2017},  biodeterioration of historical and 
   artistic objects  \cite{cappitelli_etal-b-2012,villa_etal-bs-2016}
%, obstruction of filter pores in wastewater treatment plants  \cite{nguyen_etal-mb-2012}, microbially influenced corrosion  \cite{little_etal-imr-2014}
 and fouling in food-processing equipment \citep{cappitelli_etal-fer-2014}.
The existence of a programmed generation of dispersed cells appears increasingly clear, but the challenge now is to provide the mechanistic understanding of biofilm dispersal. Thus, the principal questions that motivate this work are: 
Can we predict the colonization of new sites following biofilm dispersal?
Can we generate patterns in space and time to shed light on seeding dispersal?
%We propose a modeling approach to study how mature biofilms spread and colonize new surfaces by predicting the formation and growth of satellite colonies generated by dispersing biofilms
We propose a modeling approach to study the growth of mono-layer microbial biofilm on inert surfaces by
focusing on the biofilm spread induced by dispersal,  predicting the formation and growth of satellite colonies generated by dispersing biofilms. %migration and attachment stages.
 %%%INSERISCO LA NOTA 2 DI FEDERICA, CREDO CHE IL POSTO GIUSTO SIA QUESTO - A.T. %%%%
The importance of this work relies on the fact that the fate and behavior of dispersal cells cannot, as yet, be predicted from knowledge of the genome. Thus, a mathematical modelling of biofilm dispersion is urgently needed.
%%%% TOLGO PER INSERIRE NOTA 3 DI FEDERICA% This planar geometry is proper, for example, of phototrophic biofilms, 
%%whose microbial cells are organised as a single layer because of the limitation of nutrients.
The  planar geometry we focus on is proper of biofilm growth in oligotrophic environments (e.g., reverse osmosis membranes, stone monuments,
 surgical gauze, contact lenses, water supply pipes), where nutrient constraints limit microbial growth to thin mono-layered biofilms. 
The growth of this biofilm is characterized by two main phenomena:
the biomass expansion due to the growth of primary existing colonies,  
and the formation of new colonies due to the attachment of dispersed cells released by the primary ones, i.e., seeding dispersal.

%The proposed mechanistic model is built up as follows. Biofilm colony growth is
%modelled by using the Level Set Method \cite{sethian_etal-arfm-2003} while seeding dispersal is simulated 
%through Probability Density Function (PDF) corresponding diffusive process that governs the bacteria dispersal %behaviour

In analogy with an approach originally introduced for turbulent premixed combustion \cite{pagnini_etal-prl-2011} 
and wild-land fire propagation \cite{pagnini_etal-nhess-2014,kaur_etal-cnsns-2016},
the proposed mechanistic model is built up as follows. 
Biofilm colony growth is
modeled by using the Level Set Method \cite{sethian_etal-arfm-2003}, while seeding dispersal is simulated 
through the Probability Density Function (PDF) corresponding to the diffusive process that governs the bacteria dispersal behavior. 
%%%%%%%%%%%%%%%%%%%%%%%%%%%%%%%%%%%%%%%%%%%%%%%%%%%%%%%%%%%%%%%%%%%%%%%%%%%%%%%%%%%%%%%%%%
%%%% HO AGGIUNTO DUE DETTAGLI QUA, VEDI SOTTO PER LA DESCRIZIONE. ANDREA TRUCCHIA%%%%%%%%%
%%%%%%%%%%%%%%%%%%%%%%%%%%%%%%%%%%%%%%%%%%%%%%%%%%%%%%%%%%%%%%%%%%%%%%%%%%%%%%%%%%%%%%%%%%
The seeds attachment depends on their concentration and environmental resources availability, with the latter  characterized by its initial spatial distribution, and by the depletion effect  due to the presence of mature biofilm colonies.
The initial configuration of environmental resource availability can be modeled by setting a specific scenario or by using a random distribution.

%%%%%%%%%%%%%%%%%%%%%%%%%%%%%%%%%%%%%%%%%%%%%%%%%%%%%%%%%%%%%%%%%%%%%%%%%%%%%%%%%%%%%%%%
%%%%%%%%%%%%%%%%%%%%%%%%%\section{Math and Equations}%%%%%%%%%%%%%%%%%%%%%%%%%%%%%%%%%%%
%%%%%%%%%%%%%%%%%%%%%%%%%%%%%%%%%%%%%%%%%%%%%%%%%%%%%%%%%%%%%%%%%%%%%%%%%%%%%%%%%%%%%%%%

The surface of mature biofilm colony $\Omega$ is generally composed by an ensemble of biofilms spots $\Omega_i$
with $i=1, ..., n(t)$ where the total number $n$ depends on time $t$ because of  merging and birth of colonies.
Let $\varphi: \domain \times [0,+\infty[\, \rightarrow \R$
be a function defined on the domain of interest $\domain \subseteq \R^2$ such that the iso-line $\varphi \left(\bx,t\right) = c$
describes the evolution the boundaries of $\Omega_i$, i.e., the evolution of the colonies fronts.
Then the motion of the fronts of biofilm colonies is determined by the Level Set Equation:
%\be
%\Gamma (t) = \left\lbrace  \bx \in \mathcal D   : \varphi(x) = c                    \right\rbrace  \text{  for a fixed } c \in \mathbb R
%\ee
%\be
%\bf n = \frac{  \nabla \varphi ( \bx , t)   }{  \left\lVert {  \nabla \varphi (\bx , t)  }  \right\rVert }
%\ee
\be
\frac{  \partial \varphi(\bx ,t )} {\partial t } = u (\bx, t) \left \lVert   \nabla \varphi (\bx , t)  \right \rVert \,.
\label{LSMeqs}
\ee
In the following, the outward normal velocity $u(\bx,t)$ is assumed constant, i.e., $u(\bx, t)= u$.
 
Let the mature colonies be able to release a sufficient large number of cells whose dispersion is characterised by a random motion.
%Let $\obx_0$ be a point of the interface at the initial instant $t=0$, and
Let $\bX^\omega(t,\obx)$ be the $\omega$-realization of the trajectory of a dispersed cell with an average position $\obx=\obx(t)$ 
and initially located in $\obx(0)=\obx_0$, such that $\bX^\omega(0,\obx) = \obx_0$.
Cell trajectories are described by the one-particle density function $p^\omega(\bx;t) = \delta\left( \bx - \bX^\omega\left(t,\obx\right)\right)$,
where $\delta\left(\bx\right)$ is the Dirac $\delta$-function.
%Let $N_\omega$ be the number of the independent realizations indexed by $\omega$,
%denoting by $\la \cdot \ra$ the \textit{ensemble average}, then
%the PDF of the seeding dispersal  $B_0$ is written as
%%
%\begin{eqnarray}
%	\pd\left(\bx;t\, |\, \obx_0\right) =  \nonumber  \\ \frac{1}{N_\omega} \sum_{\omega=1}^{N_\omega}
%\delta\left( \bx - \bX^\omega\left(t, \obx_0\right) \right) =  \nonumber   \nonumber \\
%\la \delta\left( \bx - \bX^\omega\left(t, \obx_0\right) \right) \ra \,,
%\end{eqnarray}
%%
%that can be understood as the PDF counterpart of the {\it empirical} distribution function \cite{waterman_etal-ijmest-1978}.
%
%If the density of particles embodying the interface is assumed to be constant during the interface evolution,
%then an incompressibility-like condition can be stated from the Jacobian $J$ of
%the transformation $\obx \equiv \obx(t,\obx_0)$: $J=d\obx_0/d\obx = 1$.
Moreover, let the regions $\Omega$ occupied by the colonies  be conveniently marked by an indicator function $\IO(\bx,t)$.
Then, an \textit{effective indicator} $\varphi_e$, $\varphi_e (\bx, t) : \domain \times [0,+\infty[\, \rightarrow [0,1]$, 
of the region surrounded by a random front is obtained
by using the sifting property of the $\delta$-function and by averaging the  indicator function:
\begin{eqnarray}
\label{PBeq}
  \varphi_e(\bx,t)
  &=& \la \int_{\domain}    \mathcal{I}_{\Omega} (\obx,t)      \delta(\bx - \bX^\omega(t,\obx)) \, d \obx \ra \nonumber \\
  &=& \int_{\domain}   \mathcal{I}_{\Omega} (\obx,t)    \la             \delta(\bx-\bX^\omega(t,\obx)) \ra \, d\obx \nonumber \\
  &=& \int_{\domain}   \mathcal{I}_{\Omega} (\obx,t)  p(\bx;t\, |\, \obx)\, d\obx \nonumber \\
  &=& \int_{\Omega(t)} p(\bx;t\, |\, \obx) \, d\obx \,,
\end{eqnarray}
where $p\left(\bx;t\, |\, \obx\right) = \la \delta\left(\bx-\bX^\omega\left(t,\obx\right)\right) \ra$ is the
PDF of the seeding bacteria. In this work, $p\left(\bx;t\, |\, \obx\right)$ is assumed to be Gaussian.

Function $\varphi_e(\bx, t)$ provides the probability that dispersed bacteria cells arrive
in a point $\bx$ from different sources $\Omega_i (t)$. 
However, to relate this probability of arrival to a successful formation of a new biofilm colony spot,  a  criterion associated with
a reversible/irreversible attachment  due to environmental conditions and biological time  scales is needed.
With this aim, we introduce the integral field
\be
\psi (\bx ,t ) = \int_0^t   \frac{1}{\tau(\bx, \epsilon) } \varphi_e (\bx , \epsilon) \, d \epsilon \,,
\ee
that stores the signals received from the active biofilm domain $\Omega$ in the temporal interval $[0 , t]$.
We denote by $\tau (\bx , t)$ the timescale of signal storing and it is determined by 
\be
\label{eq:tau_def}
\tau (\bx , t) =  \tau_e (\bx ,t) + V (\bx, t) \,,
\ee
where $\tau_e (\bx,t)$ represents the environmental distribution of resources in absence of biofilm and  
$V(\bx , t)$ accounts for the  resource depletion performed by the biofilm.

The feedback mechanism between $\psi$ and $\varphi$ is given by the procedure
\be
\psi(\bx, t) \geq 1 \rightarrow \IO(\bx , t ) = 1 \,,
\ee
that is: when into a considered spot a certain amount of dispersed cells have established and endured a certain amount of time 
(that accounts for the environmental availability of resources) then a new colony is generated. 
Hence, the indicator function $ \IO(\bx,t)$ results to be 
\begin{equation} \label{indicatorOmega}
  \IO(\bx,t) =
  \begin{cases}
    1 \,, & \text{if}\ \varphi \left(\bx, t\right) \leq c \,\, \rm{or} \,\, \psi(\bx, t) \geq 1\,, \quad \bx \in \Omega \,, \\
    \\
    0 \,, & \text{elsewhere} \,, \quad \bx \not\in \Omega \,.
  \end{cases}
\end{equation}
%
%The indicator function at time $t=0$, i.e. $\IO(\bx, t=0)$, is denoted in the following as  $\IOzero(\bx)$, respectively.
%\iffalse
%{\color{green}

%$\tau_e$ is modelled by accounting for a random distribution of substrate spots of fixed radius, whose centers were  chosen via
%a probability distribution; here  the  uniform distribuition %$ p_{\text env } = \mathcal{U}( 0, 1) \times ( 0, 1) $  has been selected.
%has been selected.
%\be
%\tau_e (\bx) =  \tau_{{e}_0}  - \tau_{\text{spots}} \mathcal{I}_{\Omega_{\text{spots} } }(\bx)
%\ee
%The environmental inhomogeneity amplitude  $ \tau_{\text env} = \tau_{{e}_0}  - \tau_{\text{spots}}  $
%expresses the difference in colonizing time between two places, being the first with scarcity of environmental
%resources  and the second   with plenty.

Equation \eqref{eq:tau_def} shows an interplay between the availability of resources offered
from the surrounding environment and the resource depletion performed by the growth of the biofilm colonies. 
This simple formulation of the timescale for the waiting times of free cells seeding is able to generate a plethora of patterns of biological relevance.
In the following, the term $\tau_e$ is assumed constant in time, because it represents the availability of resources 
before the action of biofilm, and this changes slower than the biofilm evolution.
The term $V(\bx , t)$ is modeled by the following Poisson problem
\begin{subequations}
\be
\alpha  \Delta V (\bx, t) = \rho_b \,,
\ee
\be
V(\bx , t) |_{\bx \in \partial \domain}  = 0 \,,
\ee
\end{subequations}
where $\rho_b$ is the bacterial density inside the colonies and $\alpha$ an absorption kinetic coefficient.  
In our case, the bacterial density inside the colony is constant, and the latter equation becomes
\be
\label{elliptic}
\alpha_*  \Delta V (\bx, t) = I_\Omega(\bx,t) \,,
\ee
where $\alpha_*$ corresponds to $\alpha$ in the rescaled setting and differs for the physical dimensions.
The dynamic governed by \eqref{elliptic} depends only on $\alpha_*$ and, in spite of its simplicity, it manages 
to represent availability of biofilm resources, determining  the temporal dynamics of seeding dispersal.

In order to prove the potentiality of the proposed approach, an experimental test case has been designed and realized.
\textit{Pseudomonas aeruginosa} strain PAO1 (MH873) was used in this study as a model system of bacterial biofilms.
 In fact, the metabolically versatile \textit{P. aeruginosa} PAO1 is an opportunistic pathogen of plants, animals,
  and humans and is ubiquitously distributed in soil and aquatic habitats. Furthermore, the bacterium is genetically 
  characterized and amenable to mutagenesis and "omics" based approaches \cite{stover_etal-n-2012,walker_etal-pp-2004}.
 The microorganism was maintained at -80$^{\circ}$C in suspensions containing 20\% glycerol and 2\% peptone,
  and was grown aerobically in Tryptic Soy Broth (TSB medium) for 15h at 30$^{\circ}$C.
Dispersion experiments were conducted by using the colony-biofilm culturing system.
 %Briefly, 5 sterile black polycarbonate filter membranes (0.22 $\rm{\mu m}$ pore size and 25mm diameter) were placed in a Petri dish containing Tryptic Soy Agar (TSA medium). Bacterial cells are trapped completely by the membrane filters having a pore size smaller than the bacterial size, while nutrients and metabolites diffuse across membranes easily.
%Filter membranes were distributed accordingly to the following scheme: one filter membranes at the center of the Petri dish, and the other four at the edge of the Petri dishes, creating a cross line design (Fig. \ref{setting}). This cross line design ensured a physical distance between each membrane filters, ranging from 2mm to 20 mm.
%Fifty $\rm{\mu l}$ of cell suspension containing $1 \times 10^8$ cells were used to inoculate the central filter membrane. The plates were incubated at 30$^{\circ}$C for 72h. Every 24h the Petri dishes were observed, and the dispersal phenomenon was documented by capturing imagines with both a camera and a stereomicroscope (magnification 12X).
 Briefly, 2 sterile black polycarbonate filter membranes (0.22 $\rm{\mu m}$ pore size and 25mm diameter) were placed in each Petri dish containing Tryptic Soy Agar (TSA medium), at a distance of 2 mm from each other. 
  Bacterial cells are trapped completely by the membrane filters having a pore size smaller than the bacterial size,
   while nutrients and metabolites diffuse across membranes easily.
Fifty $\rm{\mu l}$ of cell suspension containing $1 \times 10^8$ cells were used to inoculate the central filter membrane. 
The plates were incubated at 30$^{\circ}$C for 72h. Every 24h the Petri dishes were observed, and the dispersal phenomenon was documented by capturing imagines with both a camera and a stereomicroscope (magnification 12X).

\begin{figure}[h!]
\includegraphics[width = 0.5\textwidth , trim={0cm 3cm 3cm 0cm},clip]{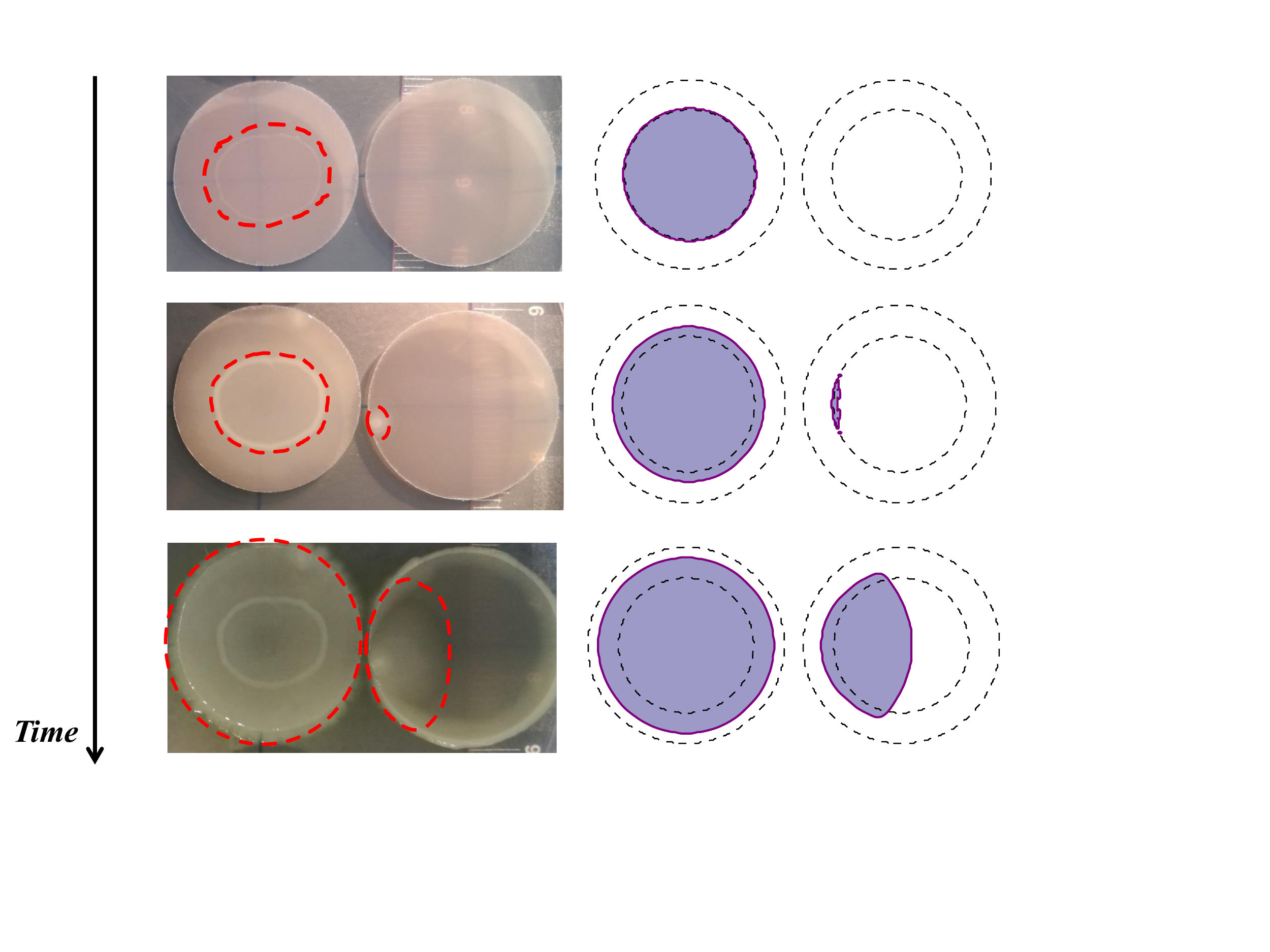}
\caption{Left column: pictures from the inoculated membrane and the  host membrane, taken at $t =  24$h, $48$h and $72$h.
The biofilm is contoured by a red dashed line. 
Right column: three stages of the numerical simulation of the experiment, where the biofilm is marked by the purple bold surface. 
%%%% UN MODO PER DIRE CHE IL NOSTRO ESPERIMENTO NON HA IL TEMPO CHE VA LINEARE DA 24 a 72! A.T. %%%
 } 	
\label{plates}
\end{figure}

Numerical solution of model \eqref{LSMeqs}--\eqref{elliptic} has been computed by setting the physical parameters as follows:
$\alpha_*=0.05 \, \rm{ms^{-2}}$, $u=10.0 \, \rm{ms^{-1}}$ inside the membranes and zero outside, and the diffusion coefficient
 of the Gaussian PDF equal to $10^3 \, \rm{ms^{-2}}$.
The numerical set-up is based on a 2D mesh $[0,220] {\rm{x}} [0,370]$ with grid step $\delta x=\delta y=1.0$.  
The numerical test concerns the two membranes: the inoculated one and seeding target (the two external dashed lines in Fig. \ref{plates}). 
These circular membranes have radius $R=60$ in grid step units and center in $(110,118)$ (the inoculated membrane) and $(110,252)$ (the target membrane).
At the initial instant, a mature biofilm colony is assumed to be present in the inoculated membrane with circular profile centered in the center of the membrane and radius $r=35$.
Furthermore, the availability of the environmental food needs to be set and it is represented by $\tau_e$ in (\ref{eq:tau_def}).
In particular, $\tau_e$ is assumed to be constant in time and ranging through a linear interpolation procedure 
from $0.01{\rm{s}}$, when $\bx$ is inside the inner disk with radius $< 0.70 R$, 
to $600.00{\rm{s}}$, when $\bx$ is outside the membranes (see the dashed circles in the right side of Fig. \ref{plates}). 
 This \textcolor[rgb]{0,0,0}{assumption} corresponds to a very large timescale for generating a new colony outside the membranes,
 which corresponds to unfavorable conditions.

The computation was done by using the facilities of BCAM by running an OpenMp-parallel finite difference  \texttt{C/Fortran} code. Its routines rely on a general-purpose library written in \texttt{Fortran2008/OpenMP}, \texttt{LSMLib} (\url{http://ktchu.serendipityresearch.org/software/lsmlib/}). The latter provides robust and efficient tools for
studying the evolution of co-dimensional fronts moving in one-, two- and three-dimensional domains. ENO algorithms are used for the sake of computing accurate space derivatives, while for the advancement in time a second order Runge--Kutta scheme was implemented.

%Outside of the two plates, the resources availability  is set very low ( by imposing $\tau_e \approx \infty$ ), therefore making the attachment outside of the two plates impossible. The two plates are modeled as having two cores of low $\tau_e$ (the two smaller, internal  dashed lines in Fig. \ref{plates}), which, in the outer ring,  augments up to the external value linearly with the radius. This is depicted by the dashed circles in Fig. \ref{plates}. The scalar velocity coefficient of the LSM equations \eqref{LSMeqs}, $u$, is set to zero outside the plates, and is set constant inside of the two plates.
 
Figure \ref{plates} shows the growth of a primary colony in the inoculated membrane and its colonization of the target membrane 
by a seeding dispersal mechanism
both for the experimental data (the pair of membranes on the left) and the proposed modeling approach (the pair  on the right).
%For this comparison, we selected the target plate where the colonization was more successful.
The Level Set Method describes the growth of the colony: first the primary one that is living in the inoculated membrane (left side membrane)
and later the secondary one in the target membrane. 
The seeding and the attachment mechanism, which are responsible for the colonization of the target membrane, are well reproduced by the model. 
In spite of the fact that the present comparison is qualitative,
it shows that the present approach is able through its modular structure to model the growth of the biomass colony  
and to take into account the different processes that simultaneously occur. 
In particular, the present approach provides a method to link a sharp interface model for the growth of biofilm \textcolor[rgb]{0,0,0}{colonies} 
and a statistical treatment for biofilm seeding. 
The modular structure allows for a detailed front propagation through a more detailed expression for the normal velocity of the colony front $u(\bx,t)$ 
and a more detailed bacterial migration through a new statical characterization.
The comparison between the experimental pictures and some frame of evolution of the model is promising
and, thanks to the modular structure,
the present approach emerges as a novel and useful method for understanding the complex dynamics displayed by microbial biofilm.

%\newpage
%{\color{red} PER GIANNI: siccome il fit e' stato fatto solo in maniera qualitativa, e le simulazioni sono state fatte adimensionali (adimensionalizzando sul diametro del Plate), non saprei come mettere  parametri. Due sono i set di parametri che ho omesso: 

%\begin{itemize}
%\item I parametri riguardanti il dx, dy, la condizione CFL,  e la griglia di calcolo in generale, la costruzione geometrica dei due Plates et similia. Vedi PDF allegato.
%\item I parametri relativi alla simulazione: \begin{itemize}
%\item $D = 1000 m/s^2$ 
%\item $\alpha_* = 0.05 m/s^2$ 
%\item  $u = 10.0 m/s$ 
%\item $\tau_e (x)$ varia tra $0.01 s $ nella parte centrale dei plates a $600.0 (\approx +\infty) s $ fuori dai plates.  Vedere il PDF allegato per ulteriori informazioni. Ho lasciato le unita' di misura per lasciare un'idea della fisica dietro i vari fenomeni.
%\end{itemize}
%\end{itemize}
%}

%trim={<left> <lower> <right> <upper>}

%\begin{figure}[h!]
%\includegraphics[width = 0.25\textwidth , trim={1cm 15cm 1cm 2cm},clip]{./images/settingstyl.pdf}
%\caption{A schematic view of the experimental setting.} 	
%\label{setting}
%\end{figure}

%%%% INSERISCO LA NOTA n.4 DI FEDERICA, METTENDO QUI IL "TO CONCLUDE" CHE HO TOLTO AL PARAGRAFO PRECEDENTE %%%%%
To conclude, we remark that one of the main motivations for studying biofilm dispersal is to provide a mechanistic model to predict 
how cells attach and proliferate to seed new biofilms. An increased understanding of the fate of dispersed cells will offer a 
broad conceptual framework for developing novel approaches to manipulate biofilm formation (either discouraging or promoting biofilm 
development) in industrial, environmental and medical applications. Thus, the ability to unravel the mechanisms of dispersal would have a great socio-economical significance, with profound implications for global health, as well as for the management of environmental microorganisms in biogeochemical cycling processes and biotechnological applications of biofilms.
Another argument supporting the significance of this model is that could be potentially applied to eukaryotes that show "biphasic" life 
cycles characterized by a dispersive phase and a sessile phase (e.g., corals, bryozoans, cancer cells). 

\smallskip
GP acknowledges the support by the Basque Government through the BERC 2014-2017 program and 
by Spanish Ministry of Economy and Competitiveness MINECO through BCAM Severo Ochoa excellence accreditation 
SEV-2013-0323 and through project MTM2016-76016-R "MIP". AT is supported by the PhD Grant "La Caixa" 2014.
LF acknowledges Progetto Giovani GNFM 2016 "Comportamenti emergenti ed auto-organizzazione in sistemi iperbolici di reazione-diffusione in ambito biologico ed ecologico." FV has received funding from the European Union Seventh Framework Programme (FP7-PEOPLE-2012-IOF) under grant agreement No. 328215.

%\bibliography{frontpropagation}% Produces the bibliography via BibTeX.

%

%\newpage
%\lipsum[1-2]
%\begin{figure}[h!]
%\includegraphics[width=\textwidth ,  trim={0cm 2cm 3cm 0cm},clip]{./images/table.pdf}
%\caption{Left column: images from the described experiment. The Red/White striped domain is the biofilm biomass.  Right column: numerical outputs from the model. The solid purple part is the  biofilm biomass.}
%\label{plates}
%\end{figure}
%\lipsum[3-10]

\end{document}